\documentclass[tighten]{revtex4-1}%
\usepackage{amsfonts}
\usepackage{amsmath}
\usepackage{amssymb}
\usepackage{graphicx}%
\begin{document}
\title{Wheels within wheels: Hamiltonian dynamics as a hierarchy of action variables}
\author{Rory J. Perkins and Paul M. Bellan}
\affiliation{Applied Physics, Caltech, Pasadena, CA 91125 }

\pacs{52.30.Cv, 52.55.Ip, 52.55.Wq}

\begin{abstract}
In systems where one coordinate undergoes periodic
oscillation, the net displacement in any other coordinate over a single period
is shown to be given by differentiation of the action
integral associated with the oscillating coordinate. This result is then
used to demonstrate that the action integral acts as a Hamiltonian for slow
coordinates providing time is scaled to the ``tick-time''of the oscillating coordinate. Numerous examples, including charged particle drifts and relativistic motion, are supplied to illustrate the varied application of these results.
\end{abstract}

\date[Date text]{date}
\maketitle
\affiliation{Applied Physics, Caltech, Pasadena, CA 91125 }
\volumeyear{year}
\eid{identifier}
\received[Received text]{date}

\revised[Revised text]{date}

\accepted[Accepted text]{date}

\published[Published text]{date}

\startpage{101}
\endpage{102}

Hamiltonian dynamics is almost ubiquitous in physics and describes such varied
phenomena as celestial mechanics, optics, fluid dynamics, quantum mechanics,
and charged particle motion in electromagnetic fields. Guiding center theory,
an approximation of Hamiltonian dynamics for charged particle  motion in
magnetic fields,   describes the motion of the  particle's cyclotron-orbit
averaged position, or guiding center~\cite{Alfven63}. \ The  guiding center can be thought of
as a \textquotedblleft quasi-particle\textquotedblright\ subject to
new types of forces and manifesting various drifts. We develop a
general model, not restricted to charged particle motion, of multi-dimensional 
systems with a periodic variable and find drifts that cannot be
calculated using guiding center theory which becomes a limited example of the
more general model. The model shows that the action integral associated with the oscillatory coordinate acts as an effective Hamiltonian for the remaining, slow
coordinates providing time is measured in clock cycles of the oscillations. We note that Hamiltonian-type aspects of action integrals have been
previously discussed in specific situations \cite{Kadomtsev1959,*Northrop1960,*White1984} but without developing a general demonstration and relying on the detailed equations of motion in their proofs.  The model presented here generates a hierarchy of \textquotedblleft
wheels-within-wheels\textquotedblright\ Hamiltonian systems such that the
action integral associated with periodic motion at any level in the hierarchy
acts as the Hamiltonian for the next slower periodic motion. \ 

Consider a 2D time-independent Hamiltonian system
$H(\xi,P_{\xi},P_{\eta})$ with an ignorable coordinate $\eta$ and where the $\xi$-motion is periodic, i.e., $\xi(t+\Delta t)=\xi(t)$, with no limit on the
amplitude of $\xi.$  $P_{\eta}$ evolves trivially: $\dot{P}_{\eta}=0$, but the $\eta$ evolution is in general nontrivial.  The period $\Delta t$ can be imagined as a clock tick over which $\eta$ undergoes a net change $\Delta\eta$.  We claim that
\begin{equation}
\Delta\eta= - \partial J / \partial P_{\eta}, \label{result}%
\end{equation}
where
\begin{equation}
J(H,P_{\eta})= \oint P_{\xi}(H,\xi,P_{\eta})d\xi\label{J}%
\end{equation}
is the action integral \cite{[][{, pg. 454, 455, 469}]Goldstein02} associated with $\xi.$ Equation (\ref{result}) means that if $J(H,P_{\eta})$ is known, the net change of
$\eta$ during one period of $\xi$ can be calculated without having to
consider the potentially complicated form of $\dot{\eta}.$

To prove Eq.~(\ref{result}), first note that there is no contribution from
differentiating the integral's bounds, so
\begin{equation}
\frac{\partial J}{\partial P_{\eta}} = \oint\frac{\partial P_{\xi
}}{\partial P_{\eta}}d\xi.\label{dJ-dPeta}%
\end{equation}
The differential of $H$ is
\begin{equation}
dH=\frac{\partial H}{\partial\xi}d\xi\ +\frac{\partial H}{\partial P_{\xi}%
}dP_{\xi}+\frac{\partial H}{\partial P_{\eta}}dP_{\eta}\ \label{dH},%
\end{equation}
so
\begin{equation}
\frac{\partial P_{\xi}}{\partial P_{\eta}}=-\frac{\partial H/\partial P_{\eta
}}{\partial H/\partial P_{\xi}}.\label{dPxi by dPeta}%
\end{equation}
Using Eq.~(\ref{dPxi by dPeta}) and Hamilton's equations in
Eq.~(\ref{dJ-dPeta}) gives
\begin{align}
\frac{\partial J}{\partial P_{\eta}} &  = - \oint\frac{\partial
H/\partial P_{\eta}}{\partial H/\partial P_{\xi}}d\xi\nonumber\\
&  = - \oint\frac{d\eta/dt\ }{d\xi/dt}d\xi=- %
\Delta\eta.\label{delta eta}%
\end{align}
If there are other ignorable coordinates in the system, then suitably adjusted
versions of Eq.~(\ref{delta eta}) apply to each of them. Equation~(\ref{result}) generalizes the well-known theorem \cite{Goldstein02} that
the period of motion is given by a partial derivative of $J$ with respect to
$H,$ namely
\begin{equation}
\Delta t = \partial J / \partial H.\label{delta t}%
\end{equation}
Equation~(\ref{delta t}) resembles Eq.~(\ref{result}) because $(t,-H)$ form a pair of canonical coordinates in extended phase space, so Eq.~(\ref{delta t}) is a special case of the theorem presented here.  The drift, or net time evolution, of $\eta$ is clearly
\begin{equation}
\frac{\Delta\eta}{\Delta t}=-\frac{\partial J/\partial P_{\eta}}{\partial
J/\partial H\ },\ \label{drift}%
\end{equation}
which generalizes the particle drifts associated with guiding
center theory.

We now relax the requirement that $\eta$ is ignorable and allow the oscillations to evolve adiabatically.  We do so by coupling the original Hamiltonian, now denoted as $H_\mathrm{loc}$, to an external system $H_\mathrm{ext}$ that is otherwise isolated.  This gives a total Hamiltonian \begin{equation}
H(\xi,P_{\xi},\eta,P_{\eta}) = H_{\mathrm{loc}}(\xi,P_{\xi},\eta,P_{\eta
})+H_{\mathrm{ext}}(\eta,P_{\eta}). \label{Htot}%
\end{equation}
We presume the system behaves as follows.  First, the $\xi$-oscillation is described by the \textit{local} Hamiltonian $H_\mathrm{loc}$ in which $\eta$ and $P_\eta$ play the role of slowly varying parameters: $d \xi / dt = \partial H_\mathrm{loc} / \partial P_\xi$ and $dP_\xi / d t = - \partial H_\mathrm{loc} / \partial \xi.$  Second, the ``parametric'' coordinates $\eta$ and $P_\eta$ are described by the \textit{total} Hamiltonian $H$, so $d \eta / d t = \partial H / \partial P_\eta$ and $d P_\eta / dt = - \partial H / \partial \eta$.  The local and external systems exchange energy, but the total energy $E=E_{\mathrm{loc}}(t) + E_{\mathrm{ext}}(t)$ is conserved since the entire system is isolated.  $J$ is defined as in Eq.~(\ref{J}) but is now also a function of $\eta$.  As in Ref.~\cite{Landau1969}, we assume it is a good approximation to hold the parametric coordinates $\eta$ and $P_\eta$ fixed while evaluating the $\xi$ action integral.  $J$ is an adibatic invariant and is thus conserved.  Furthermore, $J$ depends only on $H_{\mathrm{loc}},$ i.e. $J = J(H_\mathrm{loc}, \eta, P_\eta) = J( H - H_\mathrm{ext}(\eta, P_\eta), \eta, P_\eta),$ because $H_{\mathrm{loc}}$ is sufficient to prescribe the $\xi$ dynamics.  A proof analogous to that of Eq.~(\ref{result}) shows
\begin{equation}
\frac{\partial J}{\partial\eta}= \Delta P_{\eta},\qquad
\frac{\partial J}{\partial P_{\eta}} = - \Delta\eta. \label{delta P eta}
\end{equation}
Note that $J = J( H - H_\mathrm{ext}(\eta, P_\eta), \eta, P_\eta)$ depends on $\eta$ and $P_\eta$ both implicitly through $H_\mathrm{ext}$ and also explicitly.  Accordingly, $\Delta \eta$ and $\Delta P_\eta$ have two terms: one term comes from the explicit dependence and is the drift of the system; the second term comes from the implicit dependence and is the slow change of $H_\mathrm{ext}$.  

Equations (\ref{delta P eta}) have the makings of a Hamiltonian system with
$-J$ serving as the Hamiltonian.  They are \textit{precisely} Hamiltonian as follows.  We define discretized derivatives $d\eta/dt=\Delta\eta/\Delta t$ and
$dP_{\eta}/dt=$ $\Delta P_{\eta}/\Delta t$ that capture the net rates of change of $\eta$ and $P_{\eta}.$  Upon invocation of a rescaled time $\tau$ normalized by the $\xi$-period:
\begin{equation}
d\tau = dt / \Delta t,\ \label{def tau}%
\end{equation}
Eqs.~(\ref{delta P eta}) become
\begin{equation}
\frac{d\eta}{d\tau} = \frac{\partial}{\partial P_{\eta}}\left(  -J\right), \qquad \frac{dP_{\eta}}{d\tau} = -\frac{\partial}{\partial\eta}\left(
-J\right). \label{dPeta-dtau}%
\end{equation}
Thus, $-J$ is the Hamiltonian for the averaged system provided time is
measured in units of $\Delta t.$ It should be noted that $\tau$ is the angle
variable conjugate to $J$.

Alternatively, we can obtain a Hamiltonian for the $\xi$-averaged system by
solving $J=J(H_{loc},\eta,P_{\eta})$ for $H_{loc}=H_{loc}(J,\eta,P_{\eta})$
which upon inserting in Eq.~(9) gives%
\begin{equation}
H=H_{loc}(J,\eta,P_{\eta})+H_{ext}(\eta,P_{\eta}).\label{H averaged}%
\end{equation}
Solution of Eq.~(13)\ for $J$ gives $J=J(H,\eta,P_{\eta}).$ The differential of
$J$ using this latter form is%
\begin{equation}
dJ=\frac{\partial J}{\partial H}dH+\frac{\partial J}{\partial\eta}d\eta
+\frac{\partial J}{\partial P_{\eta}}dP_{\eta}.\label{dJ}%
\end{equation}
Equation (14)\ determines $\partial H/\partial P_{\eta}=-\left(  \partial
J/\partial P_{\eta}\right)  /\left(  \partial J/\partial H\right)  $ etc., so
using Eqs.~(10)%
\begin{align}
\frac{\partial H}{\partial P_{\eta}}  & =-\frac{\partial J/\partial P_{\eta}%
}{\partial J/\partial H}=-\frac{-\Delta\eta}{\Delta t}=\frac{d\eta}%
{dt}\label{eta dot regular time}\\
\frac{\partial H}{\partial\eta}  & =-\frac{\partial J/\partial\eta}{\partial
J/\partial H}=-\frac{\Delta P_{\eta}}{\Delta t}=-\frac{dP_{\eta}}%
{dt}.\label{P eta dot regular time}%
\end{align}
Thus, $H$, written as Eq.~(\ref{H averaged}), generates the discretized derivatives.  The term $H_\mathrm{loc}(J, \eta, P_\eta)$ is an adiabatic potential~\cite{Stefanski1985}, the residue of averaging the periodic $\xi$-motion.  For systems approximating a harmonic oscillator, $J = 2 \pi H_\mathrm{loc} / \omega(\eta, P_\eta)$, so the adiabatic potential is $H_\mathrm{loc} = J \omega / 2\pi$, showing that $J$ acts like an electrostatic charge and $\omega(\eta,P_{\eta})$ acts like an electrostatic potential.  The magnitude of this \textquotedblleft$J$-charge\textquotedblright\ depends on the amplitude of the $\xi$-oscillation.  The use of $-J(H, \eta, P_\eta)$ as a Hamiltonian with normalized time $\tau$ and the use of $H(J, \eta, P_\eta)$ with regular time are entirely equivalent.  Practically, though, there are techniques to evaluate $J$ directly~\cite{Born1960}, so using $-J$ as the Hamiltonian spares one from inverting $J$ for $H$, which might not be analytically feasible.  

%TCIMACRO{\FRAME{fhFU}{3.2301in}{2.4327in}{0pt}{\Qcb{For planar electron motion
%outside a current-carrying wire, the axial displacement $\Delta z$  can be
%derived from the radial action variable.}}{\Qlb{sample_orbit.eps}%
%}{sample_orbit.eps}{\special{ language "Scientific Word";  type "GRAPHIC";
%maintain-aspect-ratio TRUE;  display "USEDEF";  valid_file "F";
%width 3.2301in;  height 2.4327in;  depth 0pt;  original-width 5.8219in;
%original-height 4.3708in;  cropleft "0";  croptop "1";  cropright "1";
%cropbottom "0";  filename 'sample_orbit.eps';file-properties "XNPEU";}}}%
%BeginExpansion
\begin{figure}
[h]
\begin{center}
\includegraphics[
width=3.2301in
]%
{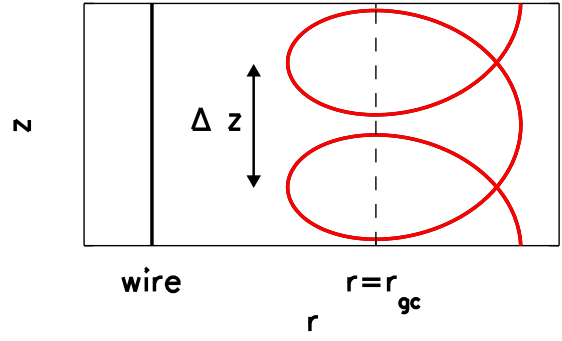}%
\caption{For planar electron motion outside a current-carrying wire, the axial
displacement $\Delta z$  can be derived from the radial action variable.}%
\label{sample_orbit.eps}%
\end{center}
\end{figure}
%EndExpansion

We now provide examples illustrating applications. Fig. \ref{sample_orbit.eps} shows an electron moving in the $rz$-plane and subjected to the magnetic field of a current-carrying wire aligned along the $z$-axis. The $z$ coordinate corresponds to $\eta$ and is ignorable; the radial motion is periodic and not ignorable because of the magnetic field gradient. The electron displaces itself an axial distance $\Delta z$ with every gyration as shown in Fig.~\ref{sample_orbit.eps}. Using the characteristic velocity $\beta=\mu_{\circ}Ie/2\pi m$ \cite{Neuberger1982},
the Hamiltonian is
\begin{equation}
H=\frac{1}{2}mv^{2}=\frac{P_{r}^{2}}{2m}+\frac{\left(  P_{z}-m\beta\ln \left(
r / R \right)  \right)  ^{2}}{2m}, \label{H-1/r}%
\end{equation}
where $R$ is an arbitrary constant of integration. $J$ can be
evaluated exactly \cite{Wouters1995} using the substitution $\cos
\theta=\left(  P_{z}-m\beta\ln\left(  r/R\right)  \right)  /mv$ and the
integral representation of the modified Bessel function $I_{n}(x)=\pi^{-1}%
\int_{0}^{\pi}e^{x\cos\theta}\cos(n\theta)d\theta$~\cite{Abramowitz1970} so
that
\begin{equation}
J = \oint P_{r}dr = 2 \pi mv r_\textrm{gc}  I_{1}\left(  \frac{v}{\beta}\right) , \label{Jplanar-1/r}%
\end{equation}
where $r_\textrm{gc} = R\exp(P_{z}/m\beta)$.  $r_\textrm{gc}$, plotted as a dashed line in Fig.~\ref{sample_orbit.eps}, is the radial position at which the $z$-velocity vanishes, as can be checked from $P_{z}=mv_{z}+m\beta\ln r/R.$  $J$ generalizes the first adiabatic invariant $\mu=mv_{\perp}^{2}/2B$~\cite{Alfven63} and reduces to $2 \pi (m/e) \mu$ when $v\ll\beta$, which for this system is the condition for the guiding center approximation to hold~\cite{Neuberger1982}. $\Delta z$ and $\Delta t$ can be computed using Eqs.~(\ref{result}) and (\ref{delta t}) and noting that $v=\sqrt{2H/m}.$ The
exact drift velocity, computed without appealing to the guiding center
approximation, is
\begin{equation}
\mathbf{v}_{d}=\frac{\Delta z}{\Delta t}\hat{z}=-v\frac{I_{1}(v/\beta)}%
{I_{0}(v/\beta)}\hat{z}.\label{wire-drift}%
\end{equation}
Equation (\ref{wire-drift}) holds for orbits of all energies even when the guiding center approximation fails. The $v\ll\beta$ limit of Eq.~(\ref{wire-drift}) reduces to the grad B drift~\cite{Bellan96} of the guiding center approximation.

Next we show how $J$ can be used as a Hamiltonian to give the magnetic mirror force~\cite{Bellan96}. For a magnetic field mainly in the $z$ direction, the cyclotron motion is essentially harmonic oscillation at the gyrofrequency $\omega = q B / m$ in the perpendicular direction, so we identify $\eta$ with $z$ and $H_{\mathrm{ext}}$ with $P_{z}^{2}/2m.$ Then
\begin{equation}
J = 2\pi \frac{H_\textrm{loc}}{\omega} = 2 \pi \frac{H-P_{z}^{2}/2m}{qB/m} = 2\pi \frac{mv_{\perp}^{2}/2}{qB/m} \label{J mu},
\end{equation}
which equals $\mu$ except for the factor $2 \pi m/q.$ If $B$
depends on $z$ then Eqs. (\ref{eta dot regular time}) and (\ref{P eta dot regular time}) become
\begin{align}
\frac{dz}{dt}  &  =-\frac{\partial J/\partial P_{z}}{\partial J/\partial
H\ }=\ \frac{P_{z}/m}{qB/m}\frac{qB}{m}=\frac{P_{z}}{m}\label{dz by dt mirror}%
\\
\frac{dP_{z}}{dt}  &  =\frac{\partial J/\partial z}{\partial J/\partial
H\ }=-\frac{qJ}%
{2\pi m}\frac{\partial B}{\partial z}=-\mu\frac{\partial B}{\partial z},
\label{dPz by dt mirror}%
\end{align}
establishing the magnetic mirror force without considering the microscopic motion.

A slightly different approach retrieves the grad B drift. Suppose
$\mathbf{B=}B_{z}(x)\hat{z},$ so $\mathbf{A}=A_{y}(x) \hat{y}$ with $B_{z}=\partial A_{y}/\partial x$.  We define the $x$ component of the guiding
center as the position $x_\textrm{gc}$ where $\ v_{y}$ vanishes: $P_y = q A_y (x_\textrm{gc})$.  Setting $\eta=y$,
Eq. (\ref{eta dot regular time}) applied to Eq.~(\ref{J mu}) gives
\begin{align}
\frac{dy}{dt}  &  =-\frac{\partial J/\partial P_{y}}{\partial J/\partial
H\ }= \mu \left( \frac{\partial B_{z}}{\partial
x}\right)_{x_\textrm{gc}} \frac{\partial x_{gc}}{\partial P_{y}}\label{y dot}.
\end{align}
We then use $\partial x_\textrm{gc} / \partial P_y = (q B(x_\textrm{gc}))^{-1}$, obtained by differentiating $P_y = q A_y (x_\textrm{gc})$ with respect to $P_{y}$. Equation (\ref{y dot}) thus becomes%
\begin{equation}
\frac{dy}{dt} = \frac{\mu}{qB_{z}} \left( \frac{\partial B_{z}}{\partial x}\right)_{x_\textrm{gc}} \label{grad B}%
\end{equation}
which is the grad $B$ drift.

\begin{figure}
[h]
\begin{center}
\includegraphics[width=.5\textwidth]{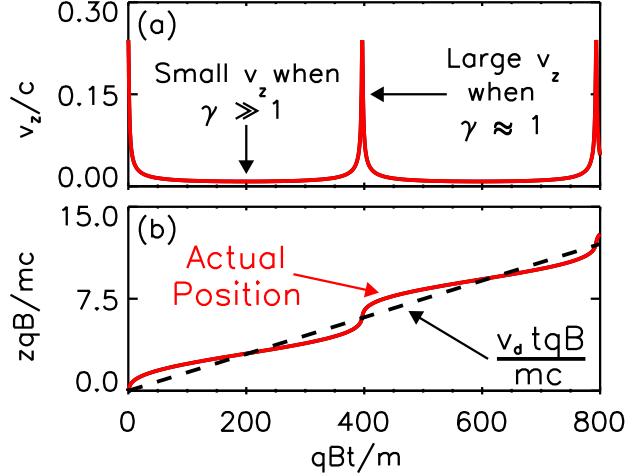}%
\caption{A particle undergoing relativistic $E \times B$ motion in the $xy$-plane with $E = 0.95 B c$ and initial momentum $P_x = 0$, $P_y = 0.7 m c$, and $P_z = 0.3 m c$.  (a) The $z$-velocity (from numerical integration of the relativistic equation of motion) is non-constant, spiking when the particle's $xy$-motion slows down so that $\gamma \approx 1$.  (b) Solid line is the numerically integrated $z$-position; dashed line, calculated using Eq.~(\ref{rel-drift}), captures the $z$-drift motion.}%
\label{fig:relativistic}%
\end{center}
\end{figure}

A surprising application arises in relativistic mechanics, where it is found
that in crossed electric and magnetic fields $\mathbf{E}%
=E\hat{x}$ and $\mathbf{B}=B\hat{z}$ with $E<Bc$ a particle's $z$-velocity is modulated by the cyclotron motion, in contrast to the non-relativistic situation where $v_{z}$ is constant and independent of the cyclotron motion.  The modulation arises from the periodic addition and subtraction of the $\mathbf{E}\times\mathbf{B}$ drift to the cyclotron velocity, which modulates $\gamma=\left(  1-v^{2}/c^{2}\right)^{-1/2}$ and hence the particle's effective mass; $v_{z}$ then varies because $v_{z}=P_{z}/\gamma m$ and  $P_{z}$ is invariant as $z$ is ignorable.  Using the relativistic canonical momenta $\mathbf{P}=m\gamma\mathbf{v}+q\mathbf{A}$ with $\mathbf{A=}Bx\hat{y}$, the electrostatic potential $\phi=-Ex,$ and the relativistic Hamiltonian $H=c\sqrt{(\mathbf{P}-q\mathbf{A})^{2}+m^{2}c^{2}%
}+q\phi,$ it is found that the relativistic $x$-action $J=\oint P_{x}dx$ evaluates to
\begin{equation}
\frac{J}{2\pi} = \frac{(BP_{y}+HE/c^{2})^{2}}{2q(B^{2}-E^{2}/c^{2})^{3/2}} + \frac{(H^{2}%
/c^{2}-P_{z}^{2}-P_{y}^{2}-m^{2}c^{2})}{2q(B^{2}-E^{2}/c^{2})^{1/2}%
}.\label{J-relativistic}%
\end{equation}
Calculating $\Delta z$ and $\Delta t$ using Eqs.~(\ref{result}) and
(\ref{delta t}) gives the $E$-dependent $z$-drift velocity
\begin{equation}
v_d = \frac{\Delta z}{\Delta t}=\frac{B^{2}-E^{2}/c^{2}}{BEP_{y}/c^{2}+B^{2}H/c^{2}%
}P_{z}.\label{rel-drift}%
\end{equation}
As shown in Fig.~(\ref{fig:relativistic}), this $v_{z}$ drift has been verified by direct numerical integration of the relativistic equation of motion $d(\gamma m\mathbf{v})/dt=q\left( \mathbf{E}+\mathbf{v\times B}\right)  \ $which shows that the modulation of $v_{z}$ is typically spiky as $\gamma \approx 1$ for a short interval during each cyclotron period and then $\gamma \gg 1$ for the remaining fraction of the cyclotron period.  Clearly, this analysis generalizes to force-drifts by replacing $\mathbf{E}$ with $\mathbf{F} / q$.

Kepler motion provides a non-relativistic and non-electromagnetic example.  The radial action is \cite{Goldstein02}
\begin{equation}
J_{r} = -2\pi \left\vert P_{\phi}\right\vert + 2 \pi \sqrt{m} mMG/ \sqrt{2 \left\vert H \right\vert },\label{J Kepler}%
\end{equation}
where $P_{\phi}$ is the conserved angular momentum. Equation~(\ref{result})
gives $\Delta\phi=\pm 2\pi$ depending on the sign of $P_{\phi}$, immediately
proving that bounded Kepler orbits are always closed.

\begin{figure}
[h]
\begin{center}
\includegraphics[width=.5\textwidth]{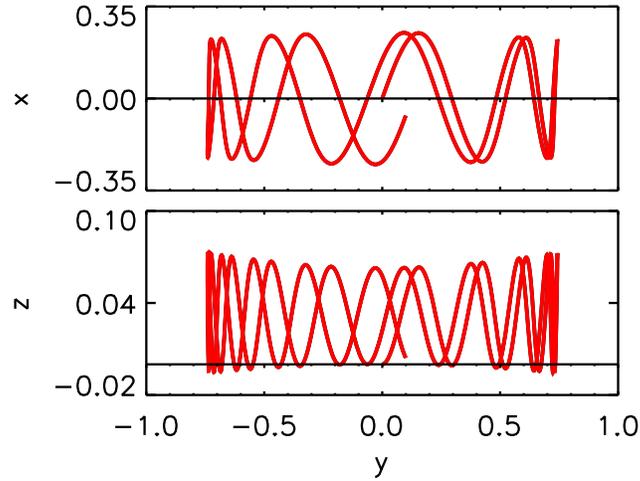}%
\caption{A particle in a thin saddle-like groove can undergo oscillatory motion due to narrowing of the groove.  $x$ is the direction across the groove, $y$ along the groove, and $z$ the vertical.  $H$ is given by Eq.~(\ref{H groove}) with $m=\kappa=\alpha=1$ and $\lambda = -.01$, and the particle starts at $x=y=0$ with $v_x=.25$ and $v_y = 1$.}%
\label{groove_sample1.jpg}%
\end{center}
\end{figure}

We now present a purely mechanical system which exhibits the equivalent of
\textquotedblleft magnetic\textquotedblright mirroring. Consider a non-relativistic particle in a long groove where the width of the groove varies with position.  The Hamiltonian is
\begin{equation}
H=\frac{P_{x}^{2}}{2m}+\frac{P_{y}^{2}}{2m}+\frac{1}{2}\kappa x^{2}\left(
1+\alpha y^{2}\right)  +\frac{\lambda}{2}y^{2};\label{H groove}%
\end{equation}
where $y$ is the distance along the groove and $x$ is the distance across the groove. Presuming that the $y$-position changes slowly relative to the oscillations across the groove (i.e., $\left\vert \alpha\right\vert $ and $\left\vert \lambda\right\vert $
are small compared to $\kappa)$ the $y$-dependent frequency of $x$%
-oscillation is
\begin{equation}
\omega(y)=\sqrt{\frac{\kappa}{m}}\sqrt{1+\alpha y^{2}}.\label{omega groove}%
\end{equation}
We identify $H_{\mathrm{loc}} = P_{x}^{2}/2m + m\omega(y)^{2}x^{2}/2,$ so the $x$-action is $J= 2 \pi H_{\mathrm{loc}}/\omega(y),$ and Eq.~(\ref{H averaged}) becomes
\begin{equation}
H=\frac{P_{y}^{2}}{2m}+\frac{\lambda}{2}y^{2} + \frac{\omega(y)}{2\pi} J.\label{H reduced groove}%
\end{equation}
Equation (\ref{P eta dot regular time}) gives an average force $\ -(J/2\pi)\,\partial\omega/\partial y=-yJ\kappa\alpha/2\pi m\omega$ in the
$y$-direction due to the increase in $x$-oscillation energy where the groove narrows.  This is a restoring force and, if sufficiently strong, can overwhelm the contribution from $\lambda.$  A negative $\lambda$ corresponds to a potential hill, and if $J=0$ the particle falls down the hill.  However, if $J$ is sufficiently large and $\alpha$ is positive, the particle does not fall down but instead oscillates about the top of the hill! This mechanical analog of a
magnetic mirror has been verified by direct numerical integration as shown in Fig.~\ref{groove_sample1.jpg}.%

%TCIMACRO{\FRAME{fhFU}{3.2361in}{2.4344in}{0pt}{\Qcb{Particle in groove.}%
%}{\Qlb{groove_sample1.jpg}}{groove_sample1.jpg}%
%{\special{ language "Scientific Word";  type "GRAPHIC";
%maintain-aspect-ratio TRUE;  display "USEDEF";  valid_file "F";
%width 3.2361in;  height 2.4344in;  depth 0pt;  original-width 5.8332in;
%original-height 4.3751in;  cropleft "0";  croptop "1";  cropright "1";
%cropbottom "0";  filename 'groove_sample1.jpg';file-properties "XNPEU";}}}%
%BeginExpansion
%EndExpansion

For oscillatory $y$-motion, Eq.~(\ref{H reduced groove}) admits an action integral in the $y$-direction, which we denote by $K$, that acts as a Hamiltonian for the $x$-averaged system.  This is a two-tier heirachy of action variables, or a wheel within a wheel.  For the reduced system, $J$ is a conserved quantity, so we develop an analog of Eq.~(\ref{result}): 
\begin{eqnarray}
\frac{\partial K}{\partial J} &=& \oint \frac{ \partial P_y (H, J, y) }{ \partial J} dy = \oint \frac{1}{\partial J / \partial P_y} dy \\
&=& \oint \frac{- 1 }{d y / d \tau} dy = - \Delta \tau, 
\end{eqnarray}
where we use Eq.~(\ref{dJ}) to evaluate $\partial P_y / \partial J$ and Eq.~(\ref{dPeta-dtau}) to evaluate $\partial J / \partial P_y$.  Since $\tau$ counts $x$-cycles, $-\partial K / \partial J$ gives the number of $x$-cycles per $y$-cycle.  If this quantity is a rational number, the trajectory is closed.  This is of interest when quantizing the system, as there is sometimes a one-to-one correspondence between periodic classical trajectories and quantum energy levels~\cite{Gutzwiller1970}.

\textit{Acknowledgments:} Supported by USDOE and NSF.

\bibliographystyle{apsrev4-1}
\bibliography{Perkins-Bellan-2010-PRL-bib-refs}

\end{document}